\documentclass[secnumarabic, twocolumn, superscriptaddress,groupedaddress, graphics,floatfix,nofootinbib, tightenlines,nobibnotes,aps,prl]{revtex4-1}

\usepackage{graphicx}  
\usepackage{dcolumn}   
\usepackage{bm}        
\usepackage{amssymb}   

\newcommand{\mnras}{{MNRAS}}

\newcommand{\apjl}{{ApJL}}
\newcommand{\aap}{{A\&A}}
\newcommand{\apjs}{{ApJS}}
\newcommand{\apss}{{Ap\&SS}}

\newcommand{\jcap}{{Journal of Cosmology and Astroparticle Physics}}

\newcommand\ringring[1]{%
  {
   \mathop{\kern0pt #1}\limits^{
     \vbox to-1.85ex{
       \kern-2ex 
       \hbox to 0pt{\hss\normalfont\kern.1em \r{}\kern-.45em \r{}\hss}%
       \vss 
     }
   }
  }
}

\begin{document}



\title{The cosmic aberration drift: proposal for a real-time detection of our acceleration through space}
\author{Julien Bel and Christian Marinoni}
\affiliation{Aix Marseille Univ, Universit\'e  de Toulon, CNRS, CPT, Marseille, France.}

\date{\today}

\begin{abstract}
Our proper acceleration with respect to the Cosmic Microwave Background results in a real-time change of the angular position of distant extragalactic sources.  
The cosmological component of this aberration drift signal, the non-inertial motion generated by the large-scale distribution of matter, can in principle be detected by future   
high-precision astrometric experiments. It  will provide interesting consistency tests  of  the standard model of cosmology,  set  independent constraints on the amplitude of the Hubble 
constant and the linear growth rate of  cosmic structures,   and  be instrumental in searching for evidence of new physics beyond the standard model. 
We present the formalism  of  this novel  cosmological test, discuss the physics to which it is sensitive and  show simulated forecasts  of the accuracy with which it can be implemented. 
\end{abstract}

\maketitle


According to the standard model of cosmology there are no special positions nor orientations in  the sky. 
At the same time, the universe exhibits a preferred (comoving) state of motion, well represented by the reference frame of the cosmic microwave background (CMB). 
The  most substantial deviation from theoretical expectations, a $\sim 10^{-3}$ dipolar modulation of the CMB temperature map,  is conventionally attributed  to our own  {\it peculiar} motion  with respect to the CMB  rest frame:
we fail to be comoving observers by about 600 km s$^{-1}$! \cite{Kogut}. 

Despite various supporting evidence \cite{Challinor, Kosow,vdipame, PlanckDop}, 
we still lack   CMB-independent measures of our speed and the ultimate proof that it is gravitationally induced by  large-scale matter fluctuations. 
The analysis of the peculiar velocity field of galaxies --- a tracer of the large-scale distribution of mass in the local universe --- is a promising but challenging approach in this direction \cite{Clarkson}. 
Current results are not conclusive and the  convergence to the CMB dipole   still disputed  in light of not yet fully understood systematics \cite{Feldman, Bilicki, Lavaux, Nusser, Feindt, Springob}.
 Several attempts   to measure our motion by searching  for dipole modulations  in high-redshift radio galaxy samples are similarly affected by observational uncertainties \cite{Baleisis}  and mostly yield velocity amplitudes larger ($> 3 \sigma$) than expected  \cite{Singal, Gibelyou, Tiwary, Colin}.

We propose to determine the nature and the amplitude of our peculiar velocity, specifically the velocity of the Local Group center (LG) with respect to the CMB, 
by constraining the rate at which it changes in time, \emph{i.e.}, by measuring our peculiar acceleration.\footnote{The observer is in gravitational free-fall
and would, of course, measure no acceleration on a local accelerometer.}
This motion adds to -- and must be disentangled from -- 
those generated by local non-cosmological dynamical processes: the acceleration of the terrestrial observer around the Sun, of the Sun around the Milky Way center \cite{Bastian, Kova2}, 
and this last around the LG (see this paper). A dimensional argument suggests that, in a uniform universe,  the physical peculiar velocity $\bm{\beta}$ of a test particle decays in time roughly as 
the inverse of the cosmic scale factor $a(t)$,  i.e. $\dot{\beta} \sim H \beta$ where $H$ is the Hubble parameter. We thus predict a cosmological aberration drift (CAD)
 effect ---  a time-dependent change of the angular position of distant (non-Galactic) sources --- 
of order $\delta \theta  \sim \dot{\beta} dt$,  where $dt $ is the time lag between  observations. Assuming that  $\beta \sim 622$ km s$^{-1}$ in the Galactic direction $l=272^\circ, b=28^\circ$ \cite{Maller2003} is  the  velocity of the LG with respect to the CMB, i.e. of the fluid element to which cosmological  perturbation theory results can be consistently applied, 
 a change in aberration of roughly  $0.4$ $\mu$as, in ten years, per each cosmic source and in some sky directions is predicted. This figure is  of  the same order of  
parallactic effects \cite{Kardashev, Quercellini}  that can in principle be extracted from all-sky, high-precision astrometric observations spaced by several years \cite{Gaianir}, and 
calls for further investigation.  Notably a detailed assessment  of the gravitational contribution  of large-scale  inhomogeneities  to the CAD signal,  as well as  of the cosmological 
information that can be inferred from our own peculiar  acceleration. 
 
 \section{ The amplitude of the Cosmological Aberration Drift}
 Consider an inhomogeneous universe  characterised by  the (longitudinal gauge)  metric 
\begin{equation}
ds^2 = e^{2\Phi(t,{\bf x})}  dt^2 -a^2(t) e^{-2 \Psi(t, {\bf x})}  \delta_{jk}dx^{j} dx^{k}
\end{equation}
where $\Phi$ and $\Psi$ are the Newtonian and curvature scalar potentials and where the smooth underlying background  has flat 
spatial hypersurfaces.  An observer $S'$,  in (time-like) geodesic motion with respect to the  comoving observer $S$, has
four-velocity $ \frac{dx^{\mu}}{d\tau}=\gamma e^{-\Phi} (1, {\bf \dot{x}})$, where $\tau$ is the proper time,  $\gamma=1/\sqrt{1-\beta^2}$ is the Lorentz boost factor, 
and where $\bm{\beta} \equiv e^{-(\Phi +\Psi)} a {\bf \dot{x}}$  is the physical  peculiar velocity  ($\dot{\,}\equiv d/dt)$.  
The aberration angle $\theta'$ between the observer's velocity and the photon arrival direction 
relates to  the un-aberrated angle $\theta$ seen by an observer at rest  as $
\cos \theta' =( \beta   +   \cos \theta)/(1+ \beta \cos \theta)$ or,  equivalently,  $   \sin \theta' = \gamma^{-1} \sin  \theta /(  1+  \beta \cos \theta )$.
The  CAD  when  $\dot{\bm \beta}$ is parallel to ${\bm \beta}$, the case of cosmological interest  (see eq. 4),   
is\footnote{The exact amplitude is 
\[
 \frac{d \theta'}{d\tau}  = 
 \frac{-\sin \theta \left[  \mathring{\beta}+\mathring{\gamma} \gamma^{-1}  \left ( \beta+ \cos \theta \right) \right] +\mathring{\theta}  \left ( 1+ \beta \cos \theta \right)}{\left[ 1+ \beta  \cos \theta  \right]^2}
\]
where $\mathring{\,}$ indicates a derivative with respect to physical time of the comoving observer $S$ ($\; \mathring{\,} \equiv e^{-\Phi}\frac{d}{dt}$).} 
\begin{equation}\label{dtpar}
\frac{d \theta'}{d\tau}   \approx    -\sin \theta' \dot{\beta}+\dot{\theta}
\end{equation}
where $\dot{\,}\equiv d/dt $.  
\begin{center}
\begin{figure}[h]
\includegraphics[width=85mm,angle=0]{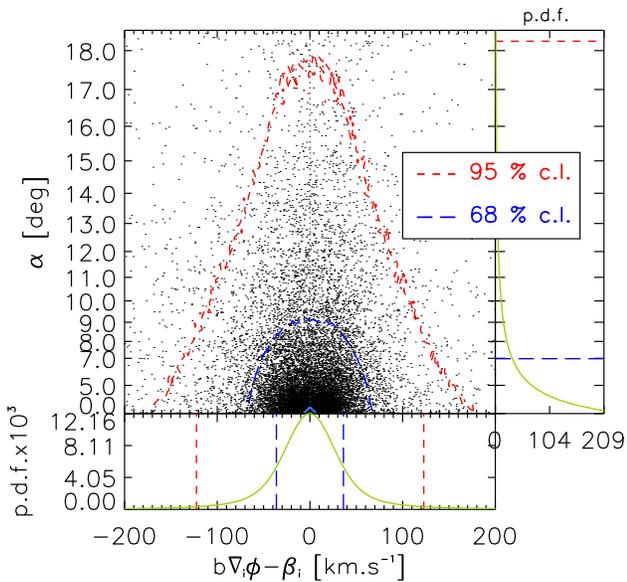}
\caption{{\it Main panel:} the angle $\alpha$ quantifies the misalignment (in degrees) between  the gradient of the gravitational potential field $\nabla \Phi$  and the direction of the peculiar  velocity vector of LG-like systems
identified in the  DEMNUni N-body simulation \cite{Castorina15}.  {\it Right panel:} the probability distribution function  (PDF) of $\alpha$ is displayed. 
The linearly reconstructed velocity vector for $68\%$   of the  LG-like systems is expected to be misaligned by at most $7^{\circ}$.
{\it Bottom panel: } PDF of the deviations between  the amplitude of the (cartesian) peculiar velocity components $\beta_i$ of simulated LG-like systems and the theoretically predicted  linear amplitude $b \nabla_i \Phi$, where $b=2f/ 3H_0 \Omega_{m0} \mu  $ (Planck $\Lambda$CDM cosmology is assumed). 
}
\label{figScatterPlot}
\end{figure}
\end{center}

The acceleration is accounted for by the equation of motion expected in the Newtonian approximation
(large sub-horizon scales) for  perfect fluids with no anisotropic stress ($\Phi=\Psi$) and negligible pressure \footnote{The acceleration of a time-like observer, in our case the LG center,  
can be evaluated exactly by solving the geodesic equation of motion  for massive   test particles. We obtain
\[
\mathring{\bm{\beta}}=-\mathcal{H} \bm{\beta}(1-\beta^2)  -  \left[ \nabla \Phi  -\bm{\beta} \cdot  \nabla(\Phi + \Psi) \bm{\beta} +\beta^2 \nabla \Psi \right] 
\] 
where  $\nabla \equiv a^{-1}e^{\Psi}\frac{\partial}{\partial {\bf x}}$ is the physical  (spatial) gradient,    and where  $ \mathcal{H} \equiv  e^{-\Phi} \left  (H-\frac{\partial \Psi}{\partial t} \right )$
is  the local Hubble factor, the log-derivative of the local scale factor ($a\,e^{-\Psi}$) with respect to the physical time 
of the comoving observer $S$ ($dT = e^\Phi dt$).} 
{\it i.e.} $\dot{\bm{\beta}}=- H \bm{\beta} -  \nabla \Phi$. The 
the amplitude of the gravitational potential follows from combining the linear continuity  equation  $\dot{\delta}+ \nabla \cdot \bm{\beta}=0$ 
for matter overdensity fluctuations $\delta$, and the Poisson equation $\Delta \Phi=4 \pi G_N \mu \bar{\rho}\delta$ 
where  $G_N$ is the Newton gravitational constant, and $\mu$, the effective gravitational constant, incorporates 
potential departure from Einstein gravity on large scales.  By assuming the existence of a factorized solution $\delta({\bf x},t)=D_+(t)\delta({\bf x})$, i.e. by
neglecting decaying modes in the evolution of the overdensity  field,  as well as rotational modes in the velocity field (compatibly with the assumption that the 
underlying perturbed metric has no vector contributions),  one finds that  the gradient of the gravitational potential 
is proportional to the peculiar velocity 
\begin{equation}
\nabla \Phi= - \frac{3 H \Omega_m}{2}\frac{\mu}{ f} \bm{\beta}
\label{bgradp}
\end{equation}
where $\Omega_m$ is the time-dependent reduced matter density of the universe and $f=d \ln D_+ / d \ln a $ is the linear growth rate of mass fluctuations.
We thus conclude that, at leading order,  proper motions evolve  as  
\begin{equation}\label{LGacce}
\dot {\bm \beta } = -H {\bm \beta} \left( 1- \frac{3 \Omega_m}{2}\frac{\mu}{f}\right). 
\label{aclin}
\end{equation}

Once reconstructed by averaging over sufficiently large spatial scales, 
the net bulk velocity ${\bm \beta}$ of distant extragalactic sources, such as quasars,  is expected to approach  zero \cite{Mak}. 
The frame defined by these distant objects thus embodies the comoving system with respect to which the LG is in relative acceleration according to eq. (\ref{LGacce}). 

\section{Sources of Noise}
 The term $\dot{\theta}$ in eq. (\ref{dtpar}) encodes contributions from the intrinsic proper motion of sources and time dependent lensing effects. This last effect, being of order $H \theta_L$ (where $\theta_L$ is the typical deflection of a distant source, $\sim 1$arcmin)  is  negligible, of order $0.004\, \mu$as yrs$^{-1}$ for distant galaxies. Proper motions,  in the most unlucky configuration, {\it i.e.} when generated by   transverse velocity as high as $\sim 1000$ km s$^{-1}$, are of order $\dot{\theta}=0.2 (Gpc/d_A)$ $\mu$as yrs$^{-1}$ where $d_A$ is the angular diameter distance \cite{syste}. Being statistically uncorrelated on large scales, as well as redshift-dependent,  they can  be  suppressed by averaging data over the azimuthal (un-aberrated) angle $\phi$, the same averaging procedure which  filters out random instrumental noise and systematic effects in identifying the positions of individual  sources. 

\begin{center}
\begin{figure}[h]
\includegraphics[width=85mm,angle=0]{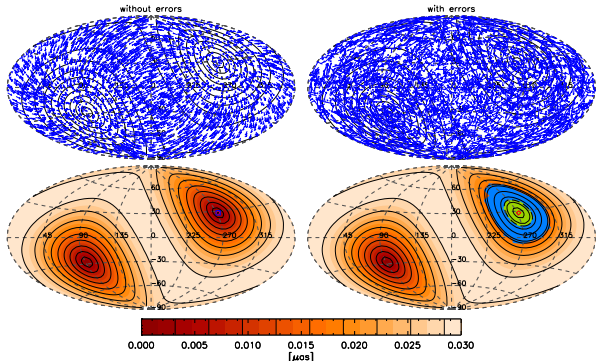}
\caption{{\it Top:} 
all-sky   isotropic  distribution of a Monte Carlo simulated sample of distant extragalactic objects.  On the left panel,  2-dimensional blue vectors show the (out of scale) CAD signal  expected for the LG moving towards the apex of the CMB temperature dipole, while on the right panel a random, and dominant,  error component,  illustrating astrometric imprecisions is added. 
{\it Bottom: }  we simulate the CAD signal reconstructed from a sample of $2 \cdot 10^6$ sources with an EoM astrometric accuracy on proper motions of $\sigma=0.6$  and $1.4 \, \mu$as yr$^{-1}$  respectively.
The red color scale shows the  amplitude of the signal (the red diamond represents the simulated direction of the observer's motion),  while the green/blue regions display the solid angle within which  68$\%$ of the  reconstructed  apex directions lie. The imprecision  in  the dipole  position is estimated using  10000 Monte Carlo realisations and  compared to the analytical predictions given in the text  (thick black lines).
} \label{figHarmonicDec}
\end{figure}
\end{center}

To extract the cosmological component of the signal one needs also to subtract  the local astrophysical motions of the terrestrial observer around the Sun, of the Sun around the Milky Way center,  and of this last  with respect to the  LG frame. While the first two effects are measurable  \cite{Kova1, Titov, TD2017} and can be disentangled (their apex is in a different direction with respect to the CAD), the motion of our own galaxy within 
LG potential must be modelled. We assume that the LG is a spherical overdensity that decoupled from the cosmological expansion and  that its gravitational potential  is essentially provided by M31, its  largest mass ($M=(1.33 \pm 0.18) \times 10^{12}M_{\odot}$), towards which the  MW -- at a distance ${\bf r}=(0.770\pm0.040)\hat{{\bf r}}$ Mpc -- is moving with a  relative velocity $\bm{\beta}_{MW/LG}=(-135 \pm  20)\hat{{\bf r}}$  km s$^{-1}$.  We obtain 
\begin{equation}
 \dot{\bm \beta}  = -H [ \bm{\beta}_{MW/LG} +\Omega_{m}/2(1+\delta_r)H_0{\bf r}]= (3\pm 48 )\hat{{\bf r}} \textit{ km s}^{-2}
 \end{equation}
where $\delta_r=16.1\pm3.9$ is the average LG spherical overdensity at the MW position. 

The two body system occupies a favorable region in configuration space, that where the Hubble expansion is exactly cancelled by the internal system gravity. Although this model  can be further refined, the contamination is expected to be one order of magnitude smaller than the cosmological signal we are looking for  (cfr. eq. (\ref{LGacce})). 

We finally investigate the accuracy to which LG-like systems satisfy the predicted  linear relation (\ref{bgradp})  between peculiar velocity ${\bm \beta}$ and  $\nabla \Phi$ describes the actual dynamics of the LG center. 
To this purpose,  we use the $z=0$  output of the DEMNUni \cite{Castorina15}, an N-body simulation of the  distribution of dark matter in the standard  $\Lambda$ Cold Dark Matter ($\Lambda$CDM) universe. 
The simulation contains $N_{cdm}=2048^3$ cold dark matter particles within a cubic periodic universe of comoving size $L=2000\;h^{-1}$Mpc. The mass resolution, $m_p = 8.27 \times 10^{10}h^{-1}M_{\odot}$, which  is an order of magnitude smaller than the mass of the Milky Way,  and the large comoving volume   make this simulation particularly well suited for analysing the galaxy clustering properties (density, velocity and acceleration fields) in the nonlinear regime.  Following \cite{Gonzalez, Maccio}, we define LG-like systems in the DEMNUni mock catalogs as regions with (top-hat) smoothed overdensity  $0<\delta_R<0.5$, where  $R=4h^{-1}$Mpc.
Figure \ref{figScatterPlot} shows that  the peculiar velocity of LG-like systems (${\bm \beta}$) is proportional to  the Newtonian gravitational force ($\nabla \Phi$)
with  a scatter which is small enough in both amplitude ($<50$ km s$^{-1}$) and direction $(<7^{\circ})$ at $68 \%\; c.l.$ (see also \cite{Nusser}).  

\section{Detection of the CAD signal and cosmological forecasts}

 The  characteristic  sinusoidal modulation of the  cosmological component of the CAD signal,  independent from frequency and distance, and, even more  
distinctively,   the fact that  its apex is predicted to be  aligned in direction  with the CMB temperature dipole,  
are essential features for gauging the reliability of real-time measurements of LG acceleration and disentangle it from other signals. The induced CAD signal, 
 being a 2D,  curl-free, vector field,  can be orthogonally decomposed on the sphere as    
$ \vec{d \theta'} = \sum V_{lm} {\bf \Psi}_{lm}$  where ${\bf \Psi}_{lm}=r \nabla Y_{lm}$
 and where  $Y_{lm}$ are the spherical harmonics.  The  $l=1$  multipole of the   divergence of the CAD vector field
$ \Theta=\nabla \cdot \vec{ d \theta'}= \sum \Theta_{lm}Y_{lm} $
is the dipole  vector $\vec{B}$ whose Cartesian coordinates,   defined through $ \Theta_{1,1} =  - \frac{1}{\sqrt{2}(B_x-iB_y)}, \;  \Theta_{10}=  B_z 
\;  \Theta_{1,-1}=  \frac{1}{\sqrt{2}(B_x+iB_y)} $ provide the apex of the observer's motion ( $ l   =\arccos{ \left [ \frac{Sg(B_y)  B_z }{ \sqrt{B_x^2+B_y^2}} \right ] }+\Pi(-B_y)\pi, \; 
 b  =\arccos{B_z/B}$,  where $Sg$  and $\Pi$ are the sign and Heaviside step functions respectively. 
%
\begin{figure*}
\includegraphics[width=48mm,angle=0]{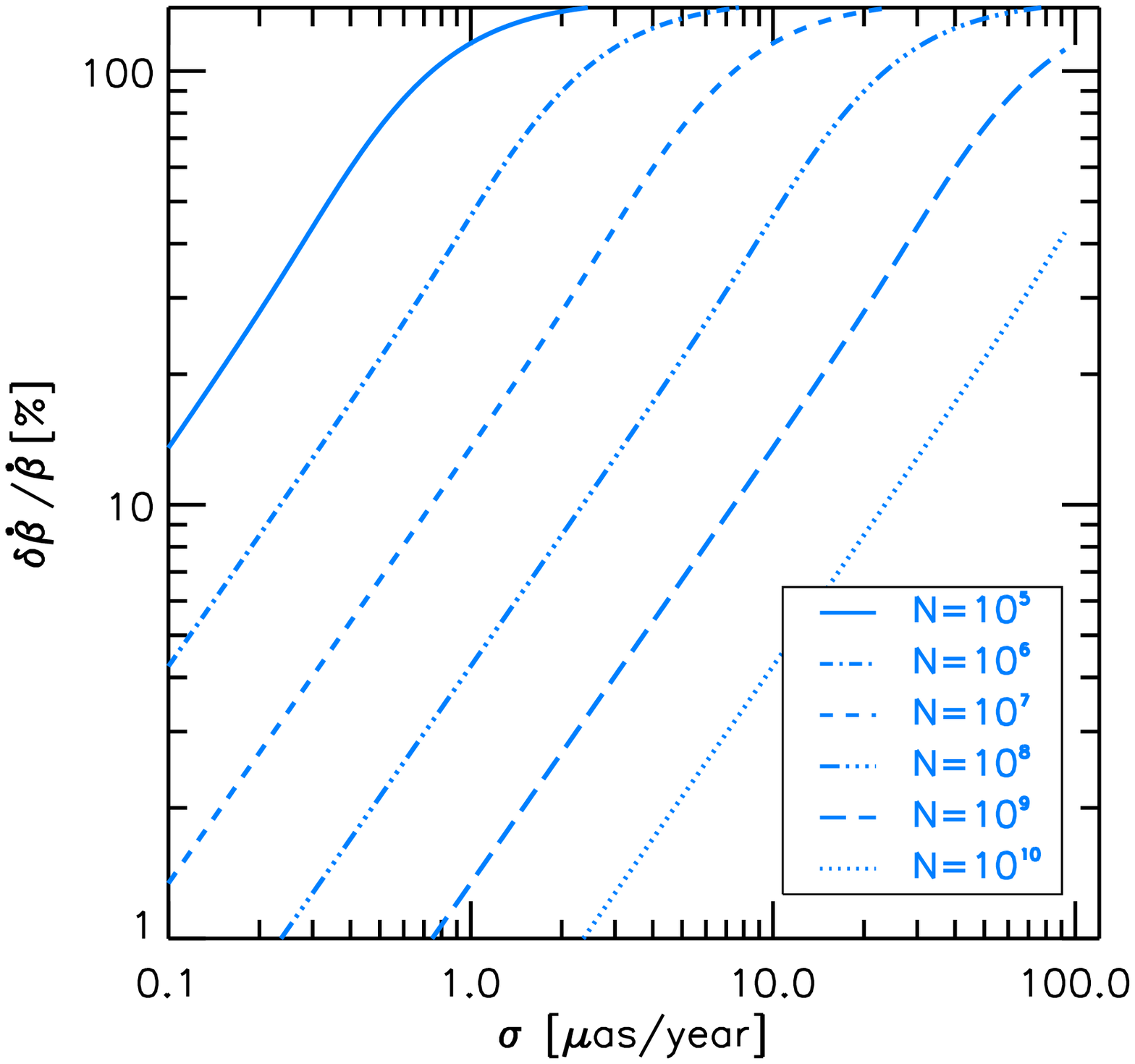}
\includegraphics[width=48mm,angle=0]{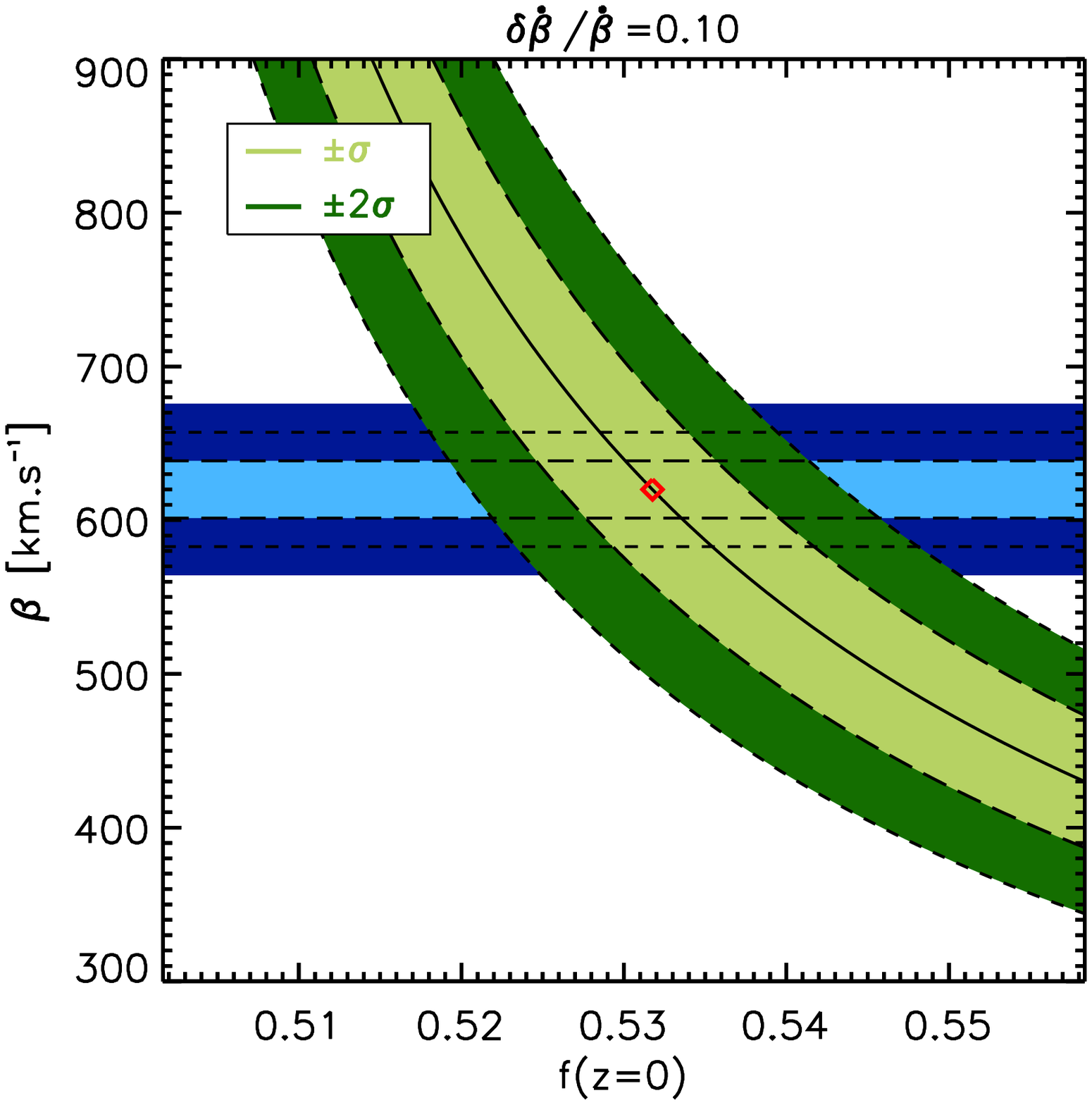}
\includegraphics[width=48mm,angle=0]{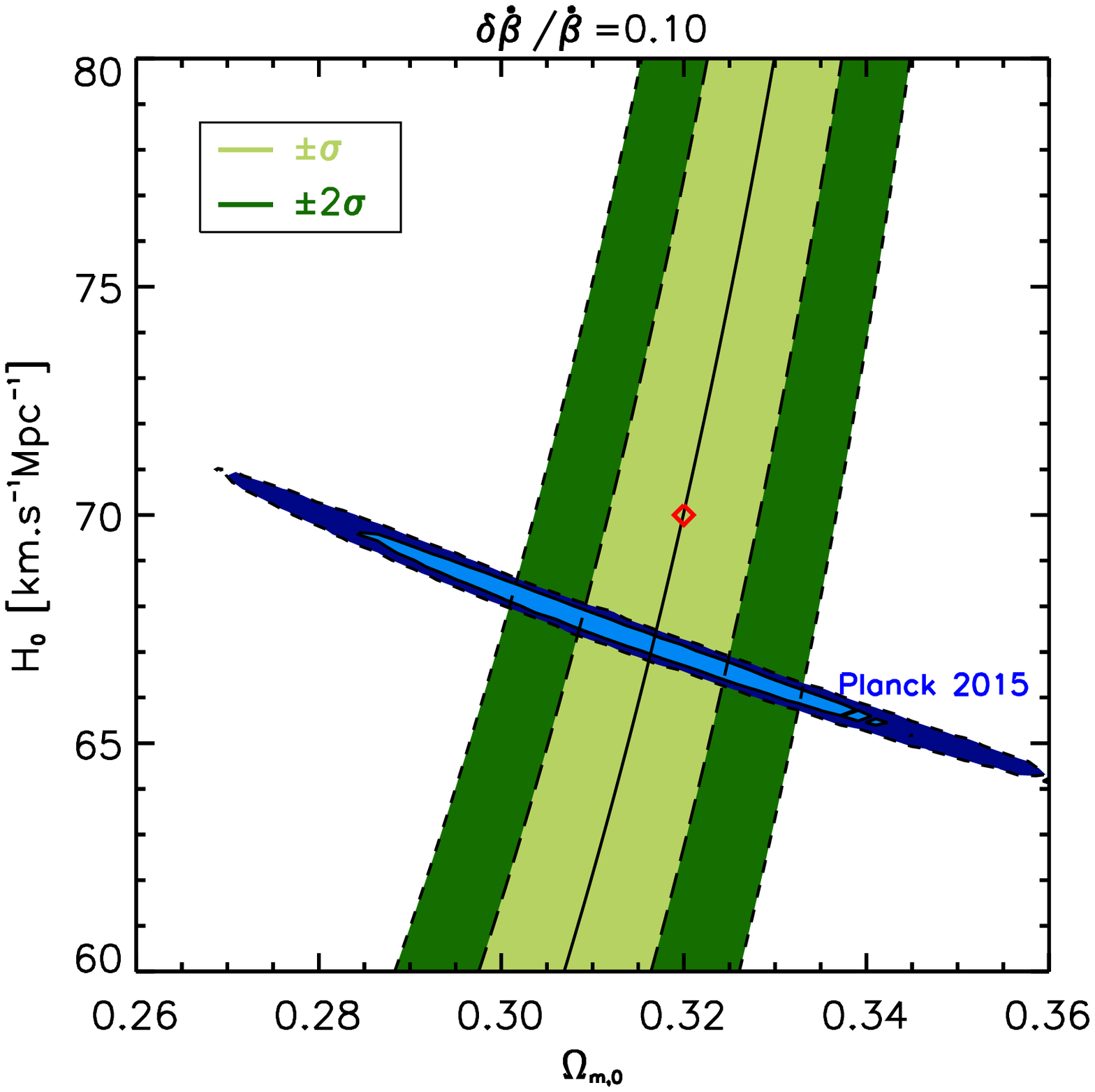}
\caption{
{\it Left: } relative precision in the estimates of the CAD signal 
as a function of the EoM astrometric error  on proper motions ($x$-axis) and of the number of cosmological sources  (as shown in the inset). We  assume
the $\Lambda$CDM cosmological model of Planck-2015.   {\it Center:}
marginalised likelihood $\mathcal{L}$ in the parameter space defined by the peculiar velocity $\beta$ and the linear growth rate $f$ at present time,
and by the Hubble constant  $H_0$ and the present-day reduced matter density $\Omega_{m0}$ ({\it right}). Confidence levels 
are drawn for $-2 \ln \mathcal{L}/\mathcal{L}_{max}=2.3$ and $6.17$.
} \label{figPrecisionConstraints}
\end{figure*}
We predict that,  in the limit in which astrometric errors dominate over statistical ones, the apex is recovered within  the solid angle 
$\Delta_\Omega = - \frac{3\pi}{N} \left (  \frac{\sigma}{ |\dot{\beta}|  }\right )^2  \ln(1-p)$ where $p$ is the desired confidence level, $\sigma $ the end-of-mission (EoM) 
proper motion error  and $N$ the number of distant sources. 
The above  analytical  predictions is confirmed  by Montecarlo simulations of the CAD signal
as shown in FIG. \ref{figHarmonicDec}. The relative error on the amplitude of the 
CAD signal, instead,  is   $\sqrt{\frac{3}{2N}}  |\dot{\beta}|^{-1}   \sigma$ and its scaling as a function of the sample size
is shown in FIG. \ref{figPrecisionConstraints}. 
  
An ongoing astrometric mission,  Gaia,  measures  proper motions for roughly $5\cdot 10^5$ QSOs as faint as  magnitude $r \sim 20$ 
with an EoM, magnitude-weighted average  proper motion error $\sigma \sim 120$ $\mu$as yr$^{-1}$ \cite{deBruijne, GAIA}.
Exploiting QSOs count statistics \cite{Ross}, we  forecast that the CAD signal could be estimated with  $20(/50)\%$ precision 
if the same sample could be targeted with an EoM proper motion error $0.7(/1.8)\mu$as yr$^{-1}$.
roughly the same accuracy with which the Gaia satellite is targeting the brightest stars of the MW.  

Encouragingly,  the necessary  astrometric precision is within  technical reach of  proposed ground experiments. 
The signal could probably be detected by analysing the $\sim  10^{10}$ extragalactic sources targeted by the  LSST \cite{LSST},  which will 
obtain proper-motion measurements of comparable accuracy to those of Gaia at its faint limit ($r<20$) and smoothly extend the error versus magnitude curve deeper by about 5 mag.
The sensitivity, field-of-view and angular resolution of the SKA will result in a  large, multi-epoch data base with precise  astrometry 
for  $\sim 2 \cdot 10^{6}$   bright compact radio sources (flux densities above $200$ $\mu$Jy).  
The forecasted proper motion accuracy of order $\sim 1$ $\mu$as yr$^{-1}$ \cite{SKA} will  allow us
to fix the amplitude of the CAD to better then $30\%$ and to locate
the  apex of the acceleration dipole  with a solid angle accuracy
$\delta\Omega/4\pi \sim 4.6\%$ (See FIG.  \ref{figHarmonicDec}).

Cosmological constraints expected  from the CAD detection with signal-to-noise ratio  $S/N=10$ are shown in FIG.  \ref{figPrecisionConstraints}. 
Besides providing CMB-independent evidence supporting  the  kinematical interpretation of the temperature dipole,  
the test will allow us to  resolve the CMB degeneracy in the estimates of   $H_0$ and $\Omega_{m0}$. 
Its sensitivity to the matter density and linear growth rate  parameters   is quite pronounced. The signal is  heavily  suppressed  if $\Omega_{m0}=0.315$,  
increasing by nearly a factor of two if  this value is lowered by $15\%$.

Interestingly,    the present day value of  $\mu/f$  predicted by modified gravity models not yet ruled out by current constraints spans a large  interval  \cite{Perenon, Salvatelli2016}.
The  resulting CAD signal  is characteristically amplified, and if  $\mu/f > 18$  could also be detected by  Gaia.  
We remark that local gravity
with $\mu/f > \frac{2}{3\Omega_{m,0}}$, which is stronger than that
predicted by $\Lambda$CDM for $\Omega_{m,0} \approx 0.3$, has a neat observational signature,  
tilting the CAD dipole  in a direction anti-aligned with respect to  that of the CMB dipole.  More generally, an eventual misalignment between the CAD apex, sensitive to the observer's proper acceleration, and the CMB apex, sensitive to the observer's velocity,  
would be a clear fingerprint of beyond standard model physics.
 
\section{Conclusions}
Precision astrometry of solar system bodies was instrumental for testing GR. We argue that  precision astrometry  (over several years)   of distant sources
will allow us  to understand the gravitational dynamics  of the LG with respect to CMB and to figure out what this local  dynamics   tell us about global properties of the universe, specifically,   its current expansion rate $H_0$ and the linear growth rate history $f$. 

The proposed method populates  the  still limited class of tests,  such as the redshift drift  \cite{Sandage, Loeb, Lake, Cora, reddist, Uzan1}, 
designed for monitoring  the real-time evolution of the universe trough  changes in cosmological observations spaced by several years. 
Unlike redshift-drift based proposals, the CAD does not necessitate any distance measurement.
It can  be implemented using  2D photometric  data only,  thus exploiting  less time-consuming observational techniques and 
imaging  instruments that are more stable, over the years, compared to spectrographs.
According to this testing scheme, an ongoing and forecasted  astrometric  missions such as Gaia could  already  provide interesting constraints on non-standard theories of gravity, notably un upper bound on  the present-day value of  $\mu/f$. 
Beyond  Gaia missions, such as LSST or SKA,   with specifically tailored  time interval between  observations,  and sufficient  astrometric precision for faint sources to the foreseeable  level predicted  in this study 
would be enough to probe  directly Gpc scales and  provide useful cosmological insights.

\bigskip
 
\acknowledgments
We are grateful to F. Piazza for insightful comments, inspiring discussions and  critical reading of the manuscript. We also  thank 
U. Abbas, R. Codur,  M.T.  Crosta,  G. Mamon  and L. Perenon for useful discussions. We acknowledge C. Carbone for kindly providing us with  the DEMNUni  $\Lambda$CDM simulation. 
We thank the referees for useful comments.
\bibliographystyle{apsrev4-1}
\bibliography{ms.bib}

\end{document}